\documentclass[12pt, draftcls, onecolumn, letterpaper]{IEEEtran}

\ifCLASSINFOpdf

\else

\fi

\ifCLASSOPTIONcompsoc
    \usepackage[caption=false, font=normalsize, labelfont=sf, textfont=sf]{subfig}
\else
    \usepackage[caption=false, font=normalsize]{subfig}
\fi
\usepackage{lipsum}%
\usepackage[dvipsnames]{xcolor}
\usepackage{algorithm,algorithmic}
\usepackage{balance}
\usepackage{multicol}   
\usepackage{cite}
\usepackage{gensymb}
\usepackage{multirow}
\usepackage{graphics}
\usepackage{epsfig}
\usepackage{graphicx}
\usepackage{epstopdf}
\usepackage{textcomp}
\usepackage{amsmath}
\usepackage{mathtools}
\usepackage{filecontents}
\usepackage{lipsum,color}
\usepackage{amssymb}
\usepackage{float}
\usepackage{colortbl} 
\usepackage{times} 
\usepackage{amsthm}  

\usepackage{amsfonts}

\theoremstyle{break}

\begin{document}
\title{Long mmWave Backhaul Connectivity Using Fixed-Wing UAVs
}

\author{\IEEEauthorblockN{ Mohammad~T.~Dabiri,~Mazen.~O.~Hasna,~Nizar~Zorba,~and~Tamer~Khattab}\\
	\IEEEauthorblockA{\textit{Department of Electrical Engineering,} 
		\textit{Qatar University}, Doha, Qatar, \\
		E-mails: (m.dabiri@qu.edu.qa; hasna@qu.edu.qa; nizarz@qu.edu.qa; tkhattab@qu.edu.qa).}
	\thanks{This publication was made possible by NPRP13S-0130-200200 from the Qatar National Research Fund (a member of The Qatar Foundation). The statements made herein are solely the responsibility of the author[s].}
}

\maketitle
\vspace{-1cm}
\begin{abstract}
This paper discusses the analysis of a fixed wing unmanned aerial vehicle (UAV)-based millimeter wave (mmWave) backhaul link, that is offered as a cost effective and easy to deploy solution  to connect a disaster or remote area to the nearest core network.
We present the optimal design of a relay system based on fixed wing UAV, taking into account the actual channel parameters such as the UAV vibrations, tracking error, real 3GPP antenna pattern, UAV's height, flight path, and the effect of physical obstacles.
%
%
The performance of the considered system is evaluated in terms of outage probability and the channel capacity, while taking into account the impact of the 
system parameters such as optimal selection of UAV flight path and antenna patterns
\end{abstract}
\IEEEpeerreviewmaketitle

\section{Introduction}

Natural disaster comprising earthquakes, hurricanes, tornadoes, floods, and other geologic processes can potentially cut or entirely destroy wired communications (e.g. fiber) infrastructure to the disaster area.
Any disruption to the fragile fiber causes data disconnections that take days to find and repair.
However, providing an alternative wireless connection link in the immediate moments after a disaster event is an essential need to facilitate rescue operations, as well as to provide internet connectivity to the people escaping from the affected area.
Providing an alternative terrestrial wireless backhaul connectivity encounters serious challenges, including creating a line of sight (LoS) between the disaster area to the nearest connected core network, especially in forest and mountainous areas.
Due to their unique capabilities such as flexibility, maneuverability,  and adaptive altitude adjustment, unmanned aerial vehicles (UAVs) acting as networked flying platforms (NFPs) can be considered as a promising solution to provide a temporary wireless backhaul connectivity while improving reliability of backhaul operations \cite{alzenad2018fso,khawaja2019survey}.
More recently, millimeter wave (mmWave) backhauling has been proposed as a promising approach for aerial communications because of three reasons \cite{dabiri2019analytical}. First, unlike terrestrial mmWave communication links that suffer from blockage, the flying nature of UAVs offers a higher probability of LoS between communication nodes. Second, the large available bandwidth at mmWave frequencies can provide high data rate point-to-point aerial communication links, as needed for the backhaul communications. Third, to compensate the negative effects of the high path-loss at the mmWave bands, the small wavelength enables the realization of a compact form of highly directive antenna arrays, which is suitable for small UAVs with limited payload.
Although NFP-based mmWave backhaul link has been studied in recent works \cite{9712177,    9714216,
	gapeyenko2018flexible,galkin2018backhaul, tafintsev2020aerial, 9411710, cicek2020backhaul, feng2018spectrum,dabiri2019analytical,dabiri20203d, yu2019uav}, the results of these studies are limited for rotary-wing UAVs. 

Rotary wing UAVs are used in cases where more maneuverability is required, for example, to provide internet service in crowded urban areas.
To keep the rotary wing UAV stable in the air, its motors are required to individually speed up or slow down its propellers, which can be time consuming, mainly due to UAV inertia. Moreover, scaling the rotary-wing UAV up to a larger size faces major challenges because more energy it needed to change speed of larger propellers. 
Another set of challenges for rotary-wing UAVs stems from restrictions on payload, altitude, and shorter flight times.
%
Being able to fly for longer times, at higher altitudes, and with heavier payloads than rotary-wing UAVs are the greatest advantages of fixed wing UAVs.
All these characteristics make them suitable for remote or disaster area applications. 
%
Based on the results of \cite{dabiri2019analytical,dabiri20203d}, to design an aerial mmWave backhaul link based on a rotary wing UAV, the goal is to find an optimal point in 3D space relative to the ground transmitter and receiver.
%
However, fixed wing UAVs cannot hover or make sharp turns, and thus, the results of the aforementioned works are not directly applicable for fixed wing UAVs.


%
\begin{figure}
	\begin{center}
		\includegraphics[width=5 in]{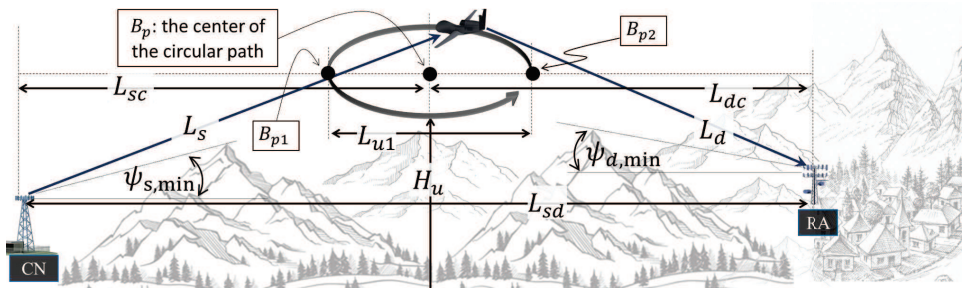}
		\caption{A fixed wing UAV acting as an NFP node in order to relay data from the nearest core network to the disaster or remote area. }
		\label{rn1}
	\end{center}
\end{figure}
%

In this study, we consider a mmWave  backhaul link based on fixed wing UAV, as shown in Fig. \ref{rn1}, that is offered as a cost effective and easy to deploy solution to connect a disaster or remote area to the nearest core network in a short time.
We fully characterize the scenario by taking into account the effects of realistic physical parameters, such as the UAV's circular path, critical points of the flight path,
heights and positions of obstacles, flight altitude, tracking errors, the severity of UAV's vibrations, the real 3D antenna pattern provided by 3GPP, atmospheric channel loss, temperature and air pressure.
We evaluate the performance of the considered system in two terms: outage probability and the average (ergodic) channel capacity.
We identify the critical points on the UAV's flight path and design the flight path of the UAV in such a way that guarantees the requested quality of service (QoS) at the considered critical points. Then, in the next step, by choosing the optimal antenna pattern, we try to maximize the capacity of the considered system.

The rest of this paper is organized as follows. We introduce the channel model of a rotary wing UAV-based mmWave backhaul link in Section II.
Our optimization problem along with the performance of the considered system, in terms of the channel capacity and the outage probability are characterized in Section III.
Using the simulation results, we study the optimal parameter design of the considered system, in Section IV.
Finally, conclusions and future road map are drawn in Section V.

\begin{table}
	\caption{The list of main notations.} 
	\centering 
	\begin{tabular}{l l} 
		\hline\hline \\[-1.2ex]
		{\bf Parameter} & {\bf Description}  \\ [.5ex] 
		\hline\hline \\[-1.2ex]
		$v\in\{t,r\}$ & This subscript is used to specify Tx and Rx antennas \\
		$q\in\{s,d\}$ & This subscript is used to specify $A_s$ and $A_d$ \\
		$w\in\{x,y\}$ & This subscript is used to specify $x$ and $y$ axes \\
		$A_{us}$ & The NFP antenna directed toward the $A_s$ \\
		$A_{ud}$ & The NFP antenna directed toward the $A_d$  \\
		$A_s$ & Antenna of CN\\
		$A_d$ & Antenna of RA\\
		$P_{t,s}$ & Transmitted power of $A_s$\\
		$P_{t,d}$ & Transmitted power of $A_{ut}$\\
		$N_{uqw}$ & Number of antenna elements of $A_{uq}$ in $w$ axis \\
		$N_{qw}$ & Number of antenna elements of $A_q$ in $w_q$ axis \\
		$\theta_{qw}$    & Instantaneous misalignment of $A_q$ in $w_q-z_q$ plane \\
		$\theta_{uqw}$    & Instantaneous misalignment of $A_{uv}$ in $w_q-z_q$ plane \\
		$\mu_{qw}$    &  Mean  of RV $\theta_{qw}$ \\
		$\sigma_{qw}^2$    &  Variance of RV $\theta_{qw}$ \\
		$\mu_{uqw}$    &  Mean of RV $\theta_{uqw}$ \\
		$\sigma_{uqw}^2$    &  Variance of RV $\theta_{uqw}$ \\
		$\lambda$ and $f_c$ &  Wavelength and carrier frequency, respectively \\
		$B_{p1}$ & The farthest and closest point to $A_d$ and $A_s$ \\
		$B_{p2}$ & The farthest and closest point to $A_s$ and $A_d$ \\
		$B_{p}$ & The center of UAV circular path \\
		%
		%
		%
		%
		$\psi_{q,\text{min}}$ & Minimum elevation angle\\
		$H_u$ & Heights of NFP \\
		$L_q$ & Link length of $A_q$ to NFP \\
		$L_{sd}$ & Horizontal distance between $A_s$ and $A_d$ \\
		$L_{qc}$ & Horizontal distance between $A_q$ and point $B_p$ \\
		%
		%
		%
		%
		\hline \hline              
	\end{tabular}
	\label{I1} 
\end{table}

\section{System Model}
We consider a fixed wing UAV acting as an NFP node in order to relay data from the nearest core network (CN) to the disaster or remote area (RA). 
The fixed-wing UAV rotates in a circular path with center $B_p$ and diameter $L_{u1}$ as depicted in Fig. \ref{rn1}.
Point $B_{p1}$ shown in Fig. \ref{rn1} is the closest and farthest point to the CN and RA, respectively. On the other hand, $B_{p2}$ is the farthest and closest point to the CN and RA, respectively.
Let $\psi_{s,\text{min}}$ and $\psi_{d,\text{min}}$ denote the minimum elevation angles of CN and RA, respectively, $L_s$ represents the link length from CN to UAV (CU), $L_d$ denotes the link length between UAV to RA (UR), $L_{sd}$ shows the distance between CN to RA, and $H_u$ stands for the UAV height.
%
%

\begin{figure}
	\begin{center}
		\includegraphics[width=5 in]{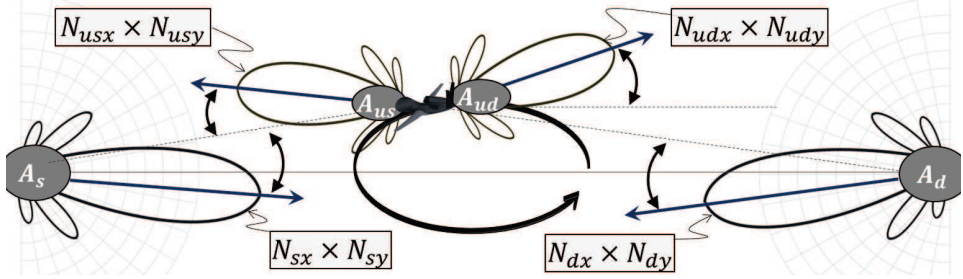}
		\caption{A graphical illustration of antenna pattern misalignment.}
		\label{fg1}
	\end{center}
\end{figure}

As shown in Fig. \ref{fg1}, our topology consists of four mmWave array antennas.
We consider that $z_s$ represents the propagation axis of CU link and axes $x_s$ and $y_s$ represent the array antenna plane perpendicular to the propagation axis. Similarly, $z_d$ axis represents the propagation axis of UR link, while axes $x_d$ and $y_d$ represent the array antenna plane perpendicular to the propagation axis $z_d$.
Let $A_s(N_{sx}\times N_{sy})$ denotes the CN array antenna composed of  $N_{sx}\times N_{sy}$ elements, where $N_{sx}$ and $N_{sy}$ are the number of antenna elements in the $x_s$ axis and $y_s$ axis directions, respectively, within the $x_s-y_s$ plane. Similarly, let $A_d(N_{dx}\times N_{dy})$ denotes the array antenna of RA node, 
$A_{us}(N_{usx}\times N_{usy})$ denotes the array antenna of the NFP directed toward the CN, and $A_{ud}(N_{udx}\times N_{udy})$ denotes the array antenna of NFP directed toward the RA, respectively. 
Antennas $A_s$ and $A_{us}$ as well as antennas $A_d$ and $A_{ud}$ try to adjust the direction of their antennas to each other.
At first, it may seem that by increasing the number of antenna elements, which leads to an increase in antenna gain, the system performance improves.
However, in practical situations, increasing the antenna gain makes the system more sensitive to antenna misalignment.
A change in the instantaneous speed and acceleration of the fixed wing UAV,
an error in the mechanical control system of UAV, mechanical noise, position estimation errors, air pressure, and wind speed can cause an alignment error between the antennas  \cite{dabiri2018channel}, as graphically illustrated in Fig. \ref{fg1}.
Therefore, the optimal design of the antenna patterns is of great importance in the presence of alignment error.
%
Let 
$\theta_{qw}\sim\mathcal{N}(\mu_{qw},\sigma_{qw}^2)$ be the instantaneous random misalignment angle of $A_q$ in $w_q-z_q$ plane, where $q\in\{s,d\}$ and $w\in\{x,y\}$.
Similarly, $\theta_{uqw}\sim\mathcal{N}(\mu_{uqw},\sigma_{uqw}^2)$ be the instantaneous random misalignment of $A_{uq}$ in $w_q-z_q$ plane.

%
%

\subsubsection{Channel Propagation Loss}
In normal atmospheric conditions, water vapor (H$_2$O) and oxygen (O$_2$) molecules are strongly absorptive of radio signals, especially at mmWave frequencies and higher.
%
%
%
%
%
%
The resulting attenuation is in excess of the reduction in radiated signal power due to free-space loss.
Channel loss (in dB) is usually expressed as
\begin{align}
	\label{f1}
	h_{L,\text{dB}}^{\text{tot}}(f_c) = 20\log\left(\frac{4 \pi L }{\lambda}\right)  + h_{L,\text{dB}}^{o,w}(f_c),
\end{align}
where $L$ is the link length (in m), $\lambda$ is the wavelength (in m), $f_c$ is mmWave frequency (in GHz),
$h_{L,\text{dB}}^{o,w}(f_c) = \frac{h_{L,\text{dB/km}}^{o,w}(f_c) L}{1000}$ is the attenuation due to oxygen and water (in dB), $h_{L,\text{dB/km}}^{o,w}(f_c)=h_{L,\text{dB/km}}^{o}(f_c)+h_{L,\text{dB/km}}^{w}(f_c)$ is the attenuation due to oxygen and water (in dB/km).
At 20°C surface temperature and at sea level, approximate expressions for the attenuation constants of oxygen and water vapor (in dB/km)  as defined by the International Telecommunications Union (ITU) are \cite{ITU_1}:
\begin{align}
	\label{po1}
	&h_{L,\text{dB/km}}^{o,0}(f_c) = 0.001\times f_c^2\\
	&\times \left\{
	\begin{array}{rl}
		\frac{6.09}{f_c^2+0.227} + \frac{4.81}{(f_c-57)^2+1.5} ~~~~~~~~~~~& ~~~  f_c<57  \\  
		h_{L,\text{dB/km}}^{o,0}(f_c=57) + 1.5(f_c-57)& ~~~  57<f_c<63 \\
		\frac{4.13}{(f_c-63)^2+1.1} + \frac{0.19}{(f_c-118.7)^2+2} ~~~~~~& ~~~  63<f_c<350
	\end{array} \right. \nonumber
\end{align}
and
\begin{align}
	\label{po2}
	&h_{L,\text{dB/km}}^{w,0}(f_c) = 0.0001\times f_c^2 \rho _0 \left(0.05  + \frac{3.6}{(f_c-22.2)^2+8.5}  \right. \nonumber \\
	& \left. + \frac{10.6}{(f_c-183.3)^2+9}    + \frac{8.9}{(f_c-325.4)^2+26.3} \right), ~~f_c<350,
\end{align}
where $\rho_0=7.5 ~\text{g/m}^3$ is the water vapor density at sea level, and $h_{L,\text{dB/km}}^{o,0}(f_c=57)$ is the value of the first expression at $f_c=57$.
In general, the attenuation constants of oxygen and water vapor are functions of altitude, since they depend on factors such as temperature and pressure. These quantities are often assumed to vary exponentially with height $H$, as $\rho(H) = \rho_0 \exp\left(-H/H_\text{scale}\right)$ where $H_\text{scale}$ is known as the scale height, which is typically 1-2 km.
From this, the specific attenuation as a function of height can be approximately modeled as
\begin{align}
	\label{po3}
	h_{L,\text{dB/km}}^{o,w}(f_c,H) = h_{L,\text{dB/km}}^{o,w,0}(f_c) \exp\left(-H/H_\text{scale}\right).
\end{align}
In our system model, both CU and UR links are slant.
For a slant atmospheric path from height $H_1$ to $H_2$ at an angle $\psi$, the total atmospheric attenuation is obtained by integration from \eqref{po3} as
\begin{align}
	\label{po4}
	h_{L,\text{dB/km}}^{o,w}(f_c) \simeq 
	\frac{h_{L,\text{dB/km}}^{o,w,0}(f_c) \left( e^{-H_1/H_\text{scale}} - e^{-H_2/H_\text{scale}} \right) H_s}{\sin(\psi)}.
\end{align}
%


\section{Performance Analysis}
For a given region with physical parameters such as air pressure, temperature, $\psi_{s,\text{min}}$, $\psi_{d,\text{min}}$, and $H_{sd}$, our aim is to adjust the tunable system parameters such as $N_{sx}$, $N_{sy}$, $N_{dx}$, $N_{dy}$, $N_{usx}$, $N_{usy}$, $N_{udx}$, $N_{udy}$, $H_u$ and $L_{sc}$, to improve system performance in terms of 
average capacity and the outage probability. These two metrics are very important in the design of wireless communication systems.
Our objective is  to maximize the channel capacity with the outage probability as a constraint (it is less than a threshold, i.e., $\mathbb{P}_\text{out}<\mathbb{P}_\text{out,th}$, where $\mathbb{P}_\text{out,th}$ is determined based on the requested quality of service (QoS)).
Our optimization problem is formulated as:
\begin{subequations} \label{opt1} 
	\begin{IEEEeqnarray}{l} 
		\displaystyle \max_{ \substack{  
				N_{sx}, N_{sy}, N_{dx}, N_{dy}, \\
				N_{usx}, N_{usy}, N_{udx}, N_{udy}, \\ H_u, L_{su}     
		}  }  ~~~~~~~{\bar{\mathbb{C}}_{e2e}}\\
		~~~~~~~~~~~\textrm{s.t.}  ~~~~~~~~~~~~\mathbb{P}_\text{out}< \mathbb{P}_\text{out,tr}~~~~~~~~~ \label{e3}\\
		~~~~~~~~~~~~~~~~~ ~~~~~~~~~H_u > H_{u,\text{min}}, \label{e2}
	\end{IEEEeqnarray}
\end{subequations}
where $\bar{\mathbb{C}}_{e2e}$ is the average channel capacity during the UAV flight time.
%
Constraint \eqref{e2} is used in order to guarantee the UAV be in the LoS of both CN and RA throughout the entire flight path. Therefore, the minimum height of the UAV should be
\begin{align}
	\label{ds1}
	&H_{u,\text{min}}= \\
	&\max\left\{(L_{sc}+\frac{L_{u1}}{2}) \sin(\psi_{s,\text{min}}),~(L_{dc}+\frac{L_{u1}}{2}) \sin(\psi_{d,\text{min}})\right\}, \nonumber
\end{align}
to ensure it satisfies the LoS for both links.
In \eqref{ds1}, we have
$L_{sc}=\sqrt{L_{s,\text{min}}^2-H_u^2}+\frac{L_{u1}}{2}$, 
$L_{dc}=\sqrt{L_{d,\text{min}}^2-H_u^2}+\frac{L_{u1}}{2}$, where $L_{s,\text{min}}$ is the link length between CN and $B_{p1}$, while $L_{d,\text{min}}$ is the link length between RA to $B_{p2}$.
We consider that the points $B_{p}$, $B_{p1}$, and $B_{p2}$ are in $[x,y]=[0,0]$, $[\frac{L_{u1}}{2},0]$, and  $[-\frac{L_{u1}}{2},0]$, respectively. 
Let $\mathcal{R}_1$ indicate the path of a semicircle that starts from point $B_{p1}$ and reaches point $B_{p2}$. 
Therefore, each point on $\mathcal{R}_1$ in the $[x-y]$ plane is specified as follows
\begin{align}
	x_{u} = \frac{L_{u1}}{2}\cos(\theta_{R1}), ~~~
	y_{u} = \frac{L_{u1}}{2}\sin(\theta_{R1}), 
\end{align}
where $0<\theta_{R1}<\pi$.
From this, the average channel capacity can be formulated as
\begin{align}
	\label{df1}
	\bar{\mathbb{C}}_{e2e} = \frac{1}{\pi}\int_{\theta_{R1}=0}^\pi \min\left\{
	\mathbb{C}_{su}(\theta_{R1}),
	\mathbb{C}_{du}(\theta_{R1})
	\right\}  \text{d}\theta_{R1},
\end{align}
where $\mathbb{C}_{su}(\theta_{R1})$ and $\mathbb{C}_{du}(\theta_{R1})$ are the average channel capacities of SU and UR links conditioned on $\theta_{R1}$, respectively. For our system model, $\mathbb{C}_{qu}(\theta_{R1})$ is a function of random variables (RVs) 
$\theta_{qx}$, $\theta_{qy}$, $\theta_{uqx}$, and $\theta_{uqy}$ and can be obtained as 
\begin{align}
	\label{c4}
	&\mathbb{C}_{qu}(\theta_{R1}) = 
	\frac{1}{4\pi^2 \sigma_{qx} \sigma_{qy} \sigma_{rqx} \sigma_{rqy} }
	\int_0^{\pi/2} \int_0^{\pi/2} \int_0^{\pi/2} \int_0^{\pi/2} \nonumber \\
	&\mathbb{C}'_{qu}(\theta_{qx},\theta_{qy},\theta_{uqx},\theta_{uqy}|\theta_{R1}) \exp\left( -\frac{(\theta_{qx}-\mu_{qx})^2}{2\sigma_{qx}}  \right) \nonumber \\
	& \times
	\exp\left( -\frac{(\theta_{qy}-\mu_{qy})^2}{2\sigma_{qy}}  \right)
	\exp\left( -\frac{(\theta_{uqx}-\mu_{uqx})^2}{2\sigma_{uqx}}  \right)
	\nonumber \\
	&\times \exp\left( -\frac{(\theta_{uqy}-\mu_{uqy})^2}{2\sigma_{uqy}}  \right)
	\text{d}\theta_{qx}  \text{d}\theta_{qy}
	\text{d}\theta_{uqx}  \text{d}\theta_{uqy},
\end{align}
where
\begin{align}
	\label{c5}
	&\mathbb{C}'_{qu}(\theta_{qx},\theta_{qy},\theta_{uqx},\theta_{uqy}|\theta_{R1}) = \nonumber \\
	&~~~~~~~~~~~~~~~ \log\left(1 + \Gamma_{q}(\theta_{qx},\theta_{qy},\theta_{uqx},\theta_{uqy}|\theta_{R1} \right),
\end{align}
and $\Gamma_{q}(\theta_{qx},\theta_{qy},\theta_{uqx},\theta_{uqy}|\theta_{R1} )$ is obtained as
\begin{align}
	\label{ro}
	&\Gamma_{q}(\theta_{qx},\theta_{qy},\theta_{uqx},\theta_{uqy}|\theta_{R1} ) = \frac{P_{t,q} h_L(L_q(\theta_{R1},L_{sc},H_u)) }{\sigma_n^2}  
	\nonumber \\
	&\times G_0(N_{qx},N_{qy}) G_0(N_{uqx},N_{uqy}) G_e(\theta_{qx},\theta_{qy}) G_e(\theta_{uqx},\theta_{uqy}) \nonumber  \\
	& \times
	\left( \frac{\sin\left(\frac{N_{qx} (k d_{qx} 
			\sin\left(\theta_{qxy}\right)
			\cos\left( \tan^{-1}\left(\frac{\tan(\theta_{qy})}{\tan(\theta_{qx})}\right) \right)+\beta_{qx})}{2}\right)} 
	{N_{qx}\sin\left(\frac{k d_{qx} 
			\sin\left( \theta_{qxy} \right)
			\cos\left( \tan^{-1}\left(\frac{\tan(\theta_{qy})}{\tan(\theta_{qx})}\right) \right)+\beta_{qx}}{2}\right)}
	\right. \nonumber \\
	&\times \left. \frac{\sin\left(\frac{N_{qy} (k d_{qy} 
			\sin\left( \theta_{qxy} \right)
			\sin\left( \tan^{-1}\left(\frac{\tan(\theta_{qy})}{\tan(\theta_{qx})}\right) \right)+\beta_{qy})}{2}\right)} 
	{N_{qy}\sin\left(\frac{k d_{qy} \sin\left(\theta_{qxy} \right)
			\sin\left( \tan^{-1}\left(\frac{\tan(\theta_{qy})}{\tan(\theta_{qx})}\right) \right)+\beta_{qy}}{2}\right)}\right)^2	 \nonumber \\	
	&\times \left( \frac{\sin\left(\frac{N_{uqx} (k d_{uqx} 
			\sin\left(\theta_{uqxy}\right)
			\cos\left( \tan^{-1}\left(\frac{\tan(\theta_{uqy})}{\tan(\theta_{uqx})}\right) \right)+\beta_{uqx})}{2}\right)} 
	{N_{uqx}\sin\left(\frac{k d_{uqx} \sin\left(\theta_{uqxy}\right)
			\cos\left( \tan^{-1}\left(\frac{\tan(\theta_{uqy})}{\tan(\theta_{uqx})}\right) \right)+\beta_{uqx}}{2}\right)}
	\right. \nonumber \\
	&\times \left. \frac{\sin\left(\frac{N_{uqy} (k d_{uqy} 
			\sin\left(\theta_{uqxy}\right)
			\sin\left( \tan^{-1}\left(\frac{\tan(\theta_{uqy})}{\tan(\theta_{uqx})}\right) \right)+\beta_{uqy})}{2}\right)} 
	{N_{uqy}\sin\left(\frac{k d_{uqy} \sin\left(\theta_{uqxy}\right)
			\sin\left( \tan^{-1}\left(\frac{\tan(\theta_{uqy})}{\tan(\theta_{uqx})}\right) \right)+\beta_{uqy}}{2}\right)}\right)^2
\end{align}
In \eqref{ro}, we have
$\theta_{qxy}=\tan^{-1}\left(\sqrt{\tan^2(\theta_{qx})+\tan^2(\theta_{qy})}\right)$,
$\theta_{uqxy}=\tan^{-1}\left(\sqrt{\tan^2(\theta_{uqx})+\tan^2(\theta_{uqy})}\right)$,
and $G_e$ is the 3GPP single element radiation pattern provided in \cite{niu2015survey}.
%
%
More details about other parameters is provided in \cite{niu2015survey} and \cite[Chapter 6]{balanis2016antenna}.

Moreover, $L_s$ and $L_d$ are the functions of $H_u$, $L_{sc}$, and $\theta_{R1}$ as
\begin{align}\label{r1}
	& L_s \!\!=\! \sqrt{ \! \left(\!L_{sc}+\frac{L_{u1}}{2}\cos(\theta_{R1})\!\right)^2+\frac{L_{u1}^2}{4}\sin^2(\theta_{R1}) + H_u^2   }\nonumber \\
	& L_d \!\!=\! \sqrt{ \! \left(\!L_{dc}\!-\!\frac{L_{u1}}{2}\cos(\theta_{R1})\!\right)^2\!+\frac{L_{u1}^2}{4}\sin^2(\theta_{R1}) + H_u^2   }.	
\end{align}
We consider that the NFP use the decode-and-forward (DF) relay system. 
Outage probability of considered system conditioned on $L_s$ and $L_d$ is obtained as \cite{hasna2003outage}:
\begin{align}
	\label{e6}
	&\mathbb{P}_{\text{out}|L_s,L_d} = \text{Prob}\left\{ 
	\min\Big[ \Gamma_{s}(\theta_{sx},\theta_{sy},\theta_{usx},\theta_{usy}|L_s,L_d ), \right.\nonumber \\
	&~~~~~~~~~~~~~~\left. \Gamma_{d}(\theta_{dx},\theta_{dy},\theta_{udx},\theta_{udy}|L_s,L_d ) \Big]
	<\Gamma_\text{th}\right\},
\end{align}
where $\Gamma_\text{th}$ is the SNR threshold.
%

\section{Simulations and Optimal System Design}
By using Monte-Carlo simulations, we now study the optimal parameter design of the considered system. The values of the parameters used in the simulations are listed in Table \ref{I2}.

\begin{table}
	\caption{Parameter values for simulations.} 
	\centering 
	\begin{tabular}{| l |c || l| c|} 
		\hline\hline 
		{\bf Parameters} & {\bf Values} &
		{\bf Parameters} & {\bf Values}  \\ [.5ex] 
		\hline\hline 
		%
		$P_{t,s} $ & 1 W &
		$P_{t,d} $ & 200 mW\\ \hline
		$N_{qw}$ & 12-18 &
		$N_{uqw}$ & 6-18 \\ \hline
		$f_c$ &  70 GHz &
		%
			$\mathbb{P}_\text{out,tr}$ & $10^{-3}$\\ \hline
			$\rho_0$&   $7.5 ~\text{g/m}^3$  &
			$T$ & $20^o$C\\ \hline
			$\beta_{qw}=\beta_{uqw}$ & 0 &
			$L_{u1}$ & 3.5 km \\ \hline
			$H_\text{scale}$ &  1.5 km &  $L_{sd}$ & 19 km \\ \hline
			$\psi_{d,\text{min}}$	& $15^o$  & $\psi_{s,\text{min}}$
			& $10^o$\\ \hline
			$d_{qw}=d_{uqw}$ & $\lambda/2$  & $\sigma_{uqx}\&\sigma_{uqy}$  & $1.5^o\&0.5$ \\ \hline
			$\sigma_{qw}\&\mu_{qw}$ &  $0.5^o \& 0.3^o$  & $\mu_{uqx}\&\mu_{uqy}$  & $1.7^o \& 1^o$ \\ 
			\hline \hline              
		\end{tabular}
		\label{I2} 
	\end{table}

	\begin{figure}
		\centering
		\subfloat[] {\includegraphics[width=3 in ]{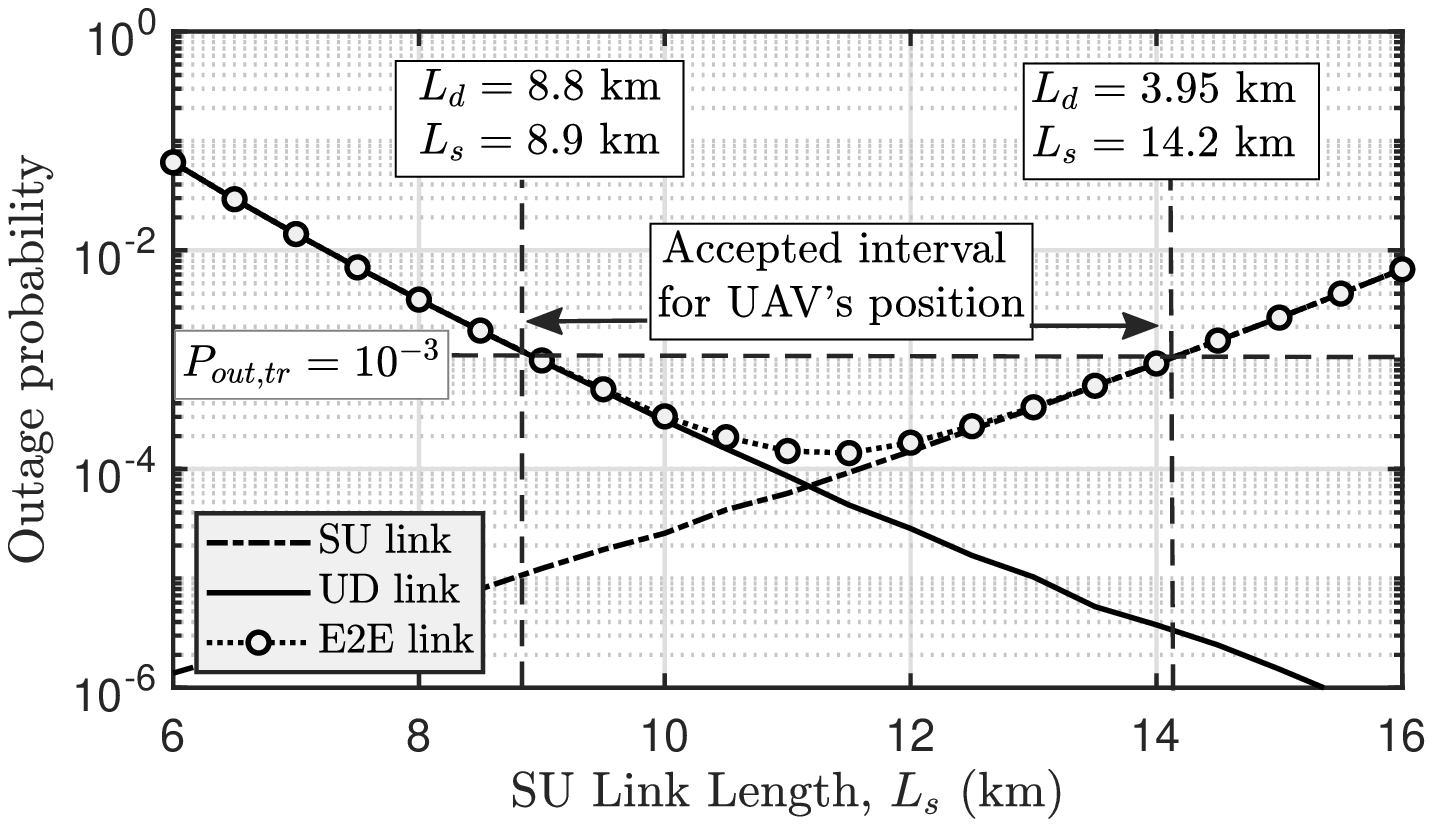}
			\label{xc1}
		}
		\hfill
		\subfloat[] {\includegraphics[width=3 in ]{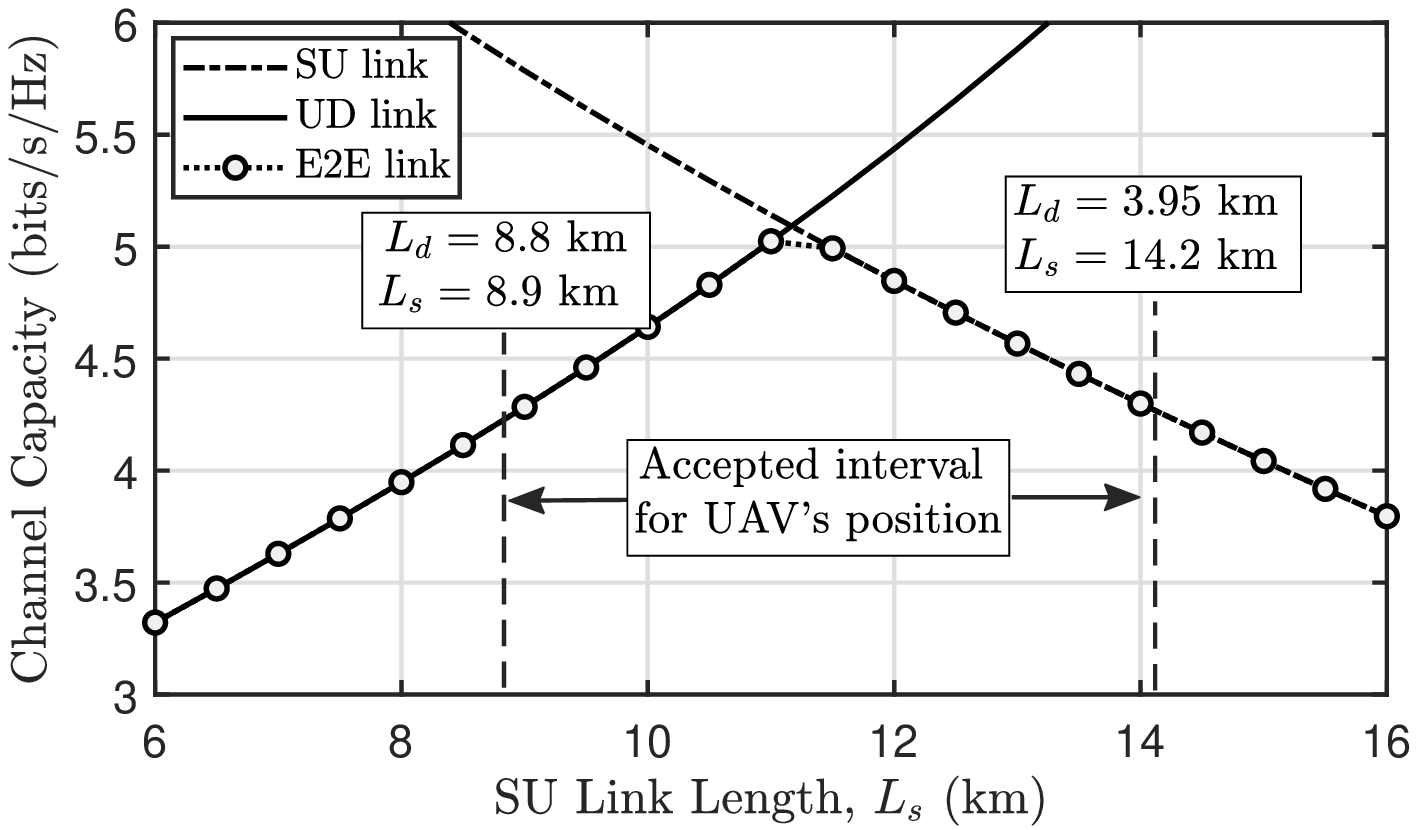}
			\label{xc2}
		}
		\caption{E2E performance of considered system versus $L_s$ and comparison with the the performance of CU and UR links in terms of (a) outage probability and (b) channel capacity. }
		\label{xc3}
	\end{figure}

	One of the important parameters is the optimal position for point $B_p$, which determines the average position of the UAV in a circular motion. As discussed, the location of point $B_p$ is adjusted in sky with the parameter $L_{sc}$. Any change in the parameters $B_p$ and $L_{sc}$ affects the values of $L_s$ and $L_d$. 
	In Fig. \ref{xc3}, the end-to-end (E2E) outage probability and channel capacity are plotted versus $L_s$ for $N_{uqx}=12$, and $N_{uqy}=N_{qw}=N_\text{max}$. 
	%
	As discussed in previous section, the E2E performance depends on the performance of CU and UR links. 
	Therefore, in Fig. \ref{xc3}, to find a better view, the performance of CU and UR links are also provided versus $L_s$.	
	The results obtained from the Fig. \ref{xc3} can be expressed in the following two remarks.
	
	{\bf Remark 1.} {\it For shorter links of $L_s$, the E2E performance can be well approximated with the performance of UR link. However, for longer links of $L_s$, E2E system performance is limited to the performance of SU link.}
	
	{\bf Remark 2.} {\it The optimal value for $L_{sc}$ is very close to the length of $L_s$ for which the capacity of SU link  is equal to the capacity of UR link.}
	
	To justify {\bf Remark 1}, note that  by increasing $L_s$, the performance of CU link decreases and at the same time $L_d$ decreases and consequently the performance of UR link improves. 
	The accepted interval for $L_s$ and $L_d$ shown in Fig. \ref{xc1} is to guarantee condition \eqref{e3}. Based on \eqref{e3} and {\bf Remark 1}, we can conclude the following remark. 
	
	{\bf Remark 3.} {\it In order to guarantee constraint \eqref{e3} along the circle flight path, it is necessary that $L_s<L_{s,\text{max}}$ and $L_d<L_{d,\text{max}}$ where $L_{s,\text{max}}$ and $L_{d,\text{max}}$ are obtained as
		\begin{align}
			\label{e5}
			&\mathbb{P}_\text{out,tr} = \text{Prob}\left\{ 
			\Gamma_{s}(\theta_{sx},\theta_{sy},\theta_{usx},\theta_{usy}|L_{s,\text{max}} ) 
			<\Gamma_\text{th}\right\}, \\
			&\mathbb{P}_\text{out,tr} = \text{Prob}\left\{ 
			\Gamma_{d}(\theta_{dx},\theta_{dy},\theta_{udx},\theta_{udy}|L_{d,\text{max}} ) 
			<\Gamma_\text{th}\right\}.
		\end{align}
	}

	\begin{figure}
		\centering
		\subfloat[] {\includegraphics[width=3 in]{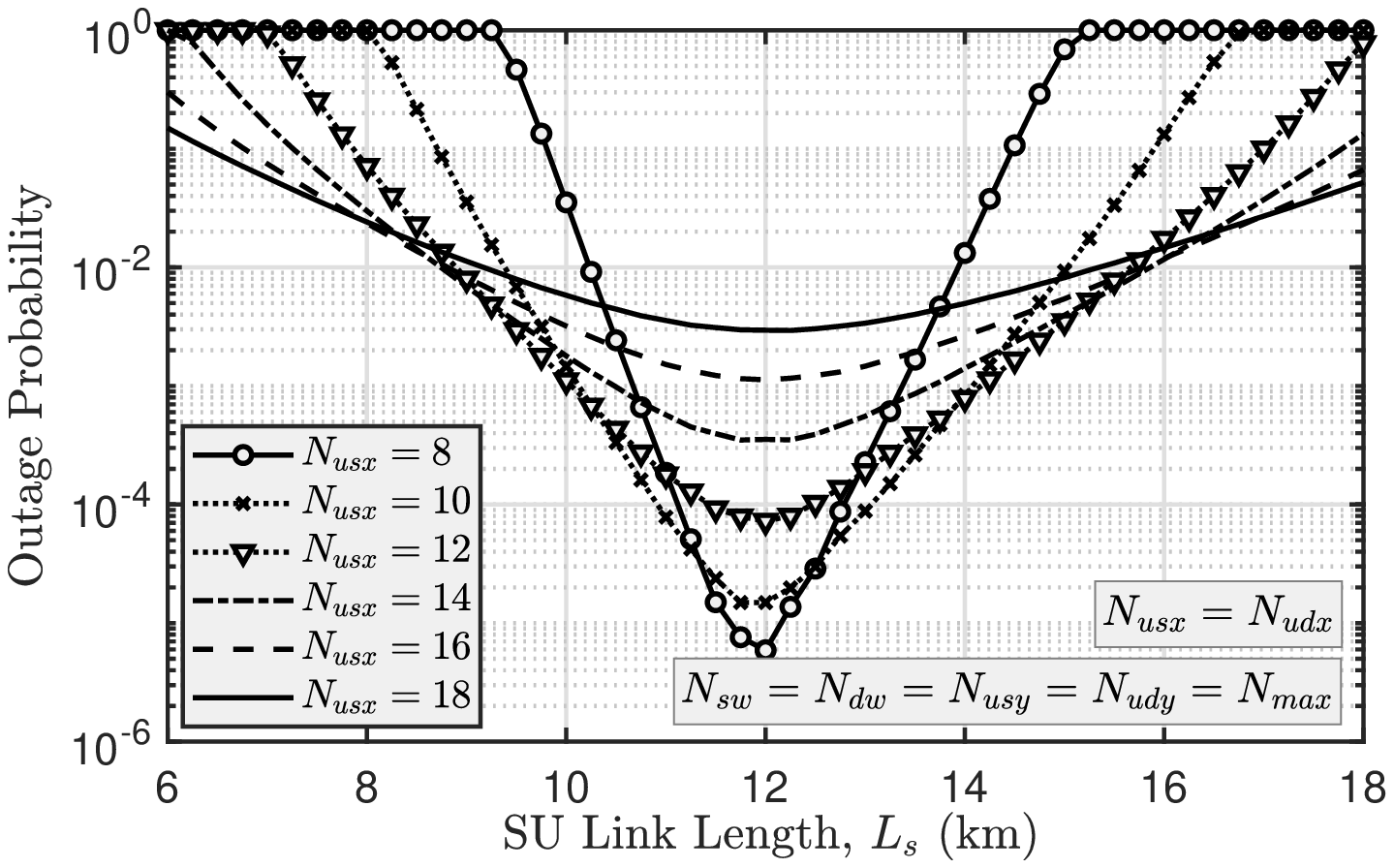}
			\label{ml1}
		}
		\hfill
		\subfloat[] {\includegraphics[width=3 in ]{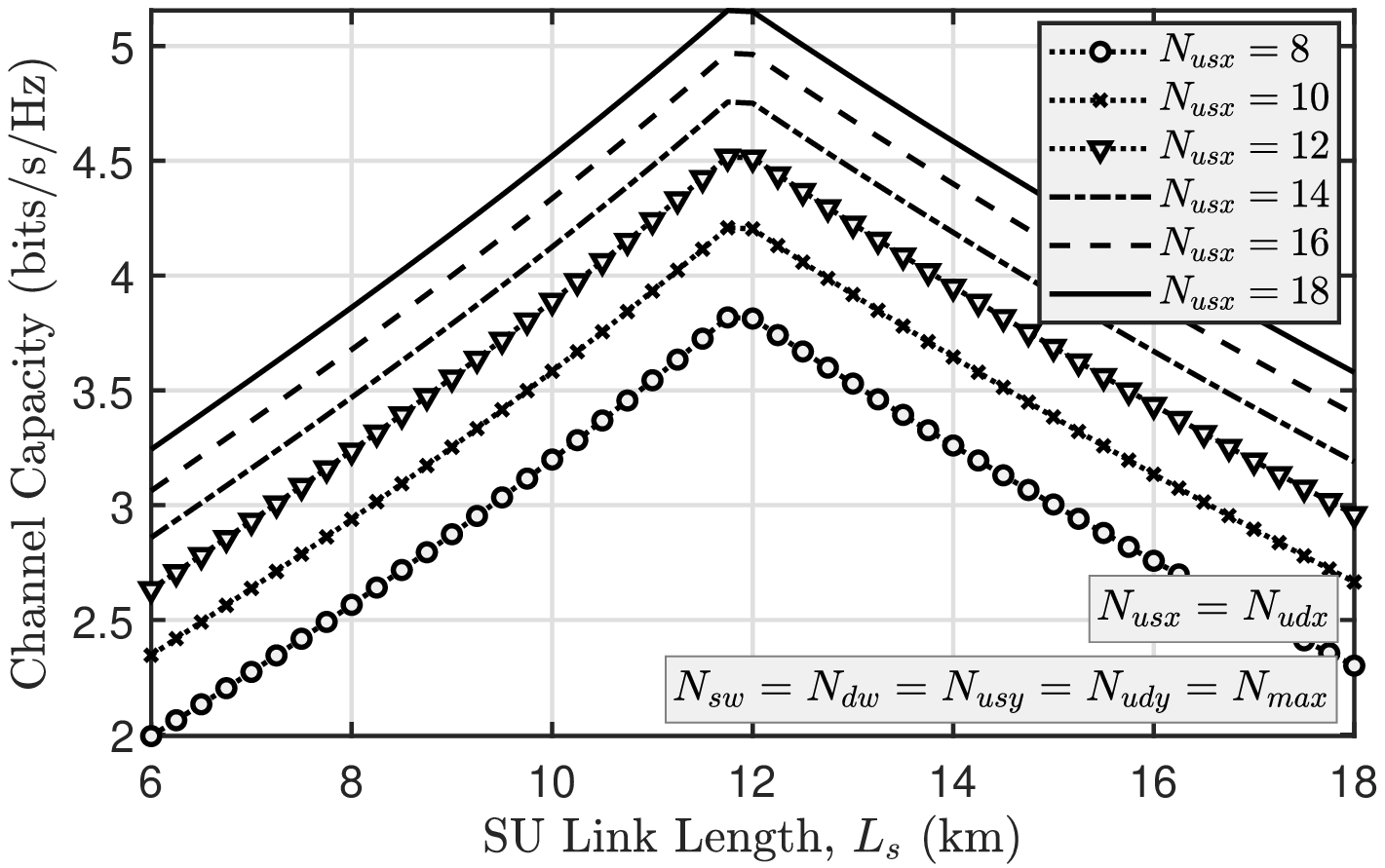}
			\label{ml2}
		}
		\caption{E2E performance of the considered system versus $L_s$ for different values of $N_{uqx}$ in terms of (a) outage probability and (b) channel capacity.}
		\label{ml3}
	\end{figure}

	\begin{figure}
		\centering
		\subfloat[] {\includegraphics[width=3 in]{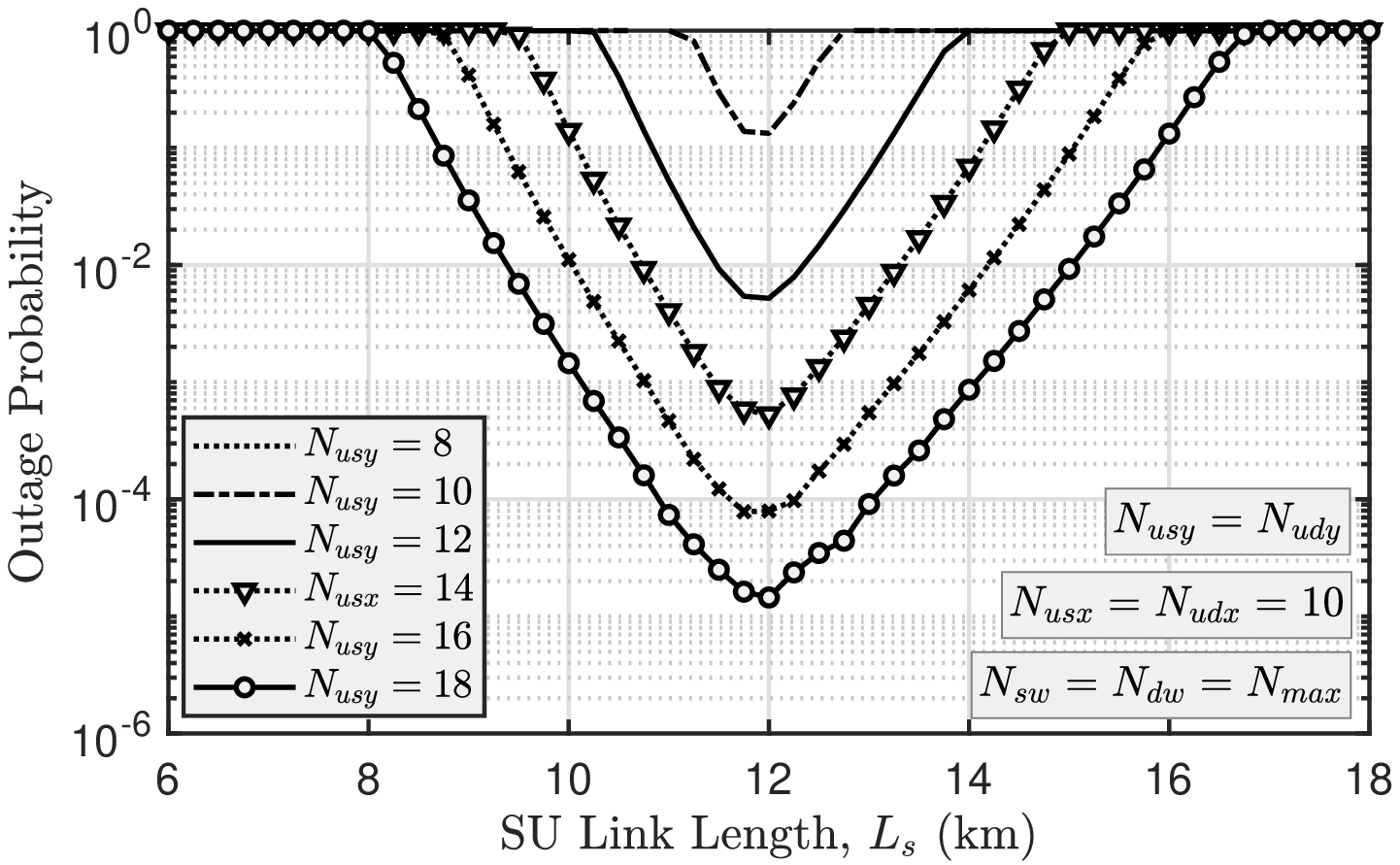}
			\label{xl1}
		}
		\hfill
		\subfloat[] {\includegraphics[width=3 in]{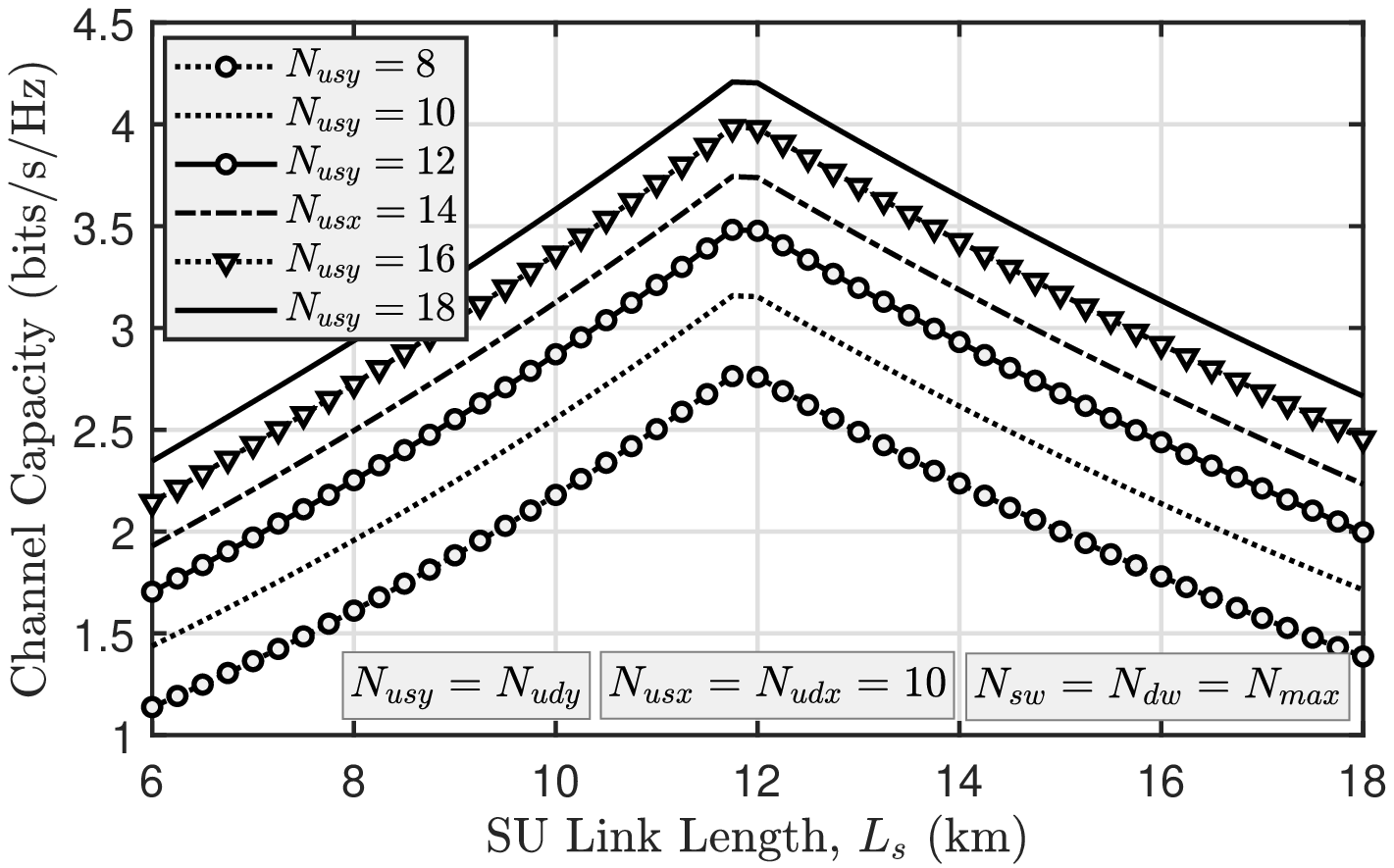}
			\label{xl2}
		}
		\caption{E2E performance of the considered system versus $L_s$ for different values of $N_{uqy}$ in terms of (a) outage probability and (b) channel capacity.}
		\label{xl3}
	\end{figure}

	Note that the parameters $L_{s,\text{max}}$ and $L_{d,\text{max}}$ are functions of the tunable parameters $N_{qw}$, $N_{uqw}$, $P_{t,q}$, frequency band, $H_u$, temperature, wind speed, UAV's instabilities, tracking system error, etc. 
	In the direction of the UAV movement, the variance of antenna misalignment are higher than the variance of antenna misalignment in direction perpendicular to the UAV movement.
	Therefore, unlike hovering UAVs, for the considered fixed-wing UAV, we have $\sigma_{uqx}>\sigma_{uqy}$, and thus, we expect the optimal number for antenna elements to be different along the $x_q$ and $y_q$ axes \cite{dabiri20203d}.
	In Figs. \ref{ml3} and \ref{xl3}, we analyze the E2E performance of the considered system versus $L_s$ for different values of $N_{uqx}$ and $N_{uqy}$, respectively. 
	From the results of Fig. \ref{ml3}, although the channel capacity gets higher by increasing $N_{uqx}$, the antenna beamdwidth decreases for large $N_{uqx}$ and the system becomes more sensitive to alignment errors. Therefore, as we observe, the channel capacity is maximized 
	for $N_{uqx}=16$ and 18. However, for these values of $N_{uqx}$, we have $\mathbb{P}_\text{out}>\mathbb{P}_\text{out,tr}$ for all values of $L_s$ 
	and therefore, the required QoS in \eqref{e3} is not guaranteed.
	Therefore, it seems that the optimal value for $N_{uqx}$ is equal to 14. However, for $N_{uqx}=14$, the accepted interval for $L_d$ (clearly illustrated and discussed in Fig. \ref{xc3}) is lower than $L_{u1}$ and thus, the considered system cannot guarantee \eqref{e3} along the entire flight path of the UAV. As a result, for the parameter values given in Table \ref{I2}, the optimal value for $N_{uqx}$, which maximizes the channel capacity and at the same time guarantees \eqref{e3} along the entire flight path, is $N_{uqx}=12$.

	Now, in Fig. \ref{xl3}, the effect of $N_{uqy}$ is investigated in the direction of $y_q$, which has a lower misalignment error.
	Unlike $N_{uqx}$, it is observed that with increasing $N_{uqy}$, the system performance in both outage probability and  channel capacity improves.
	The reason for this is that for lower misalignment errors, the antenna is less sensitive to alignment error and an antenna with a higher gain can be used to increase the channel capacity.
	For the values provided in Table \ref{I2}, the optimal value $N_{uqy}=N_\text{max}=18$. Note that in practice, due to weight and aerodynamic limitations of the UAV payload, a very large antenna cannot be used and we have to consider a maximum for $N_{quw}$. Based on the results of Figs. \ref{ml3} and \eqref{xl3}, we can conclude the following remark.
	
	{\bf Remark 4.} {\it If $\sigma_{uqy}<\sigma_{uqx}$, then the optimal value for $N_{uqy}$ will be greater than the optimal value for $N_{uqx}$.}

	Finding the optimal value for the system parameters is very important, especially for antenna patterns. 
	It should be noted that the optimal values for tunable parameters will change as a result of any variations in channel parameters. For instance,  UAV's instability changes due to wind speed variations  which, in turn, will change the alignment error severity.
	Therefore, finding and updating the optimal values for the number of antenna pattern elements is vital, and the results of {\bf Remark 4} limit the search space for discrete values of $N_{uqy}$ and $N_{uqx}$, and thus, reducing the optimization time.
	%
	\begin{figure}
		\begin{center}
			\includegraphics[width=3 in]{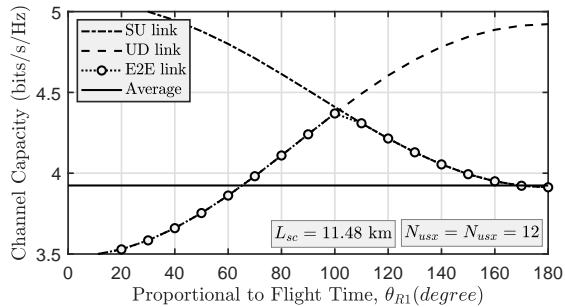}
			\caption{Channel capacity versus UAV's flight time.}
			\label{mn1}
		\end{center}
	\end{figure}
	%
	
	It is important to note that integrating on $L_s$ to achieve the average channel capacity in the entire circular motion path of the UAV is not correct because there is not a linear relationship between flight path and $L_s$.
	In Fig. \ref{mn1}, the channel capacity versus $\theta_{R1}$ which is proportional to the UAV's flight, is plotted for optimal values obtained for the positions of point $L_{sc}$ and UAV's antenna patterns. 
	Comparing the results of Fig. \ref{mn1} with the results of Figs. \ref{ml2} and \ref{xl2}, it is observed that the channel capacity distribution in proportion to the flight time is different from the capacity distribution relative to $L_s$. 
	In order to obtain the average channel capacity and find the optimal values for antenna patterns, integration on $\theta_{R1}$ must be done. As a result, how to use the results of the figures relative to $L_s$ and $\theta_{R1}$ is described in the following remark.

	{\bf Remark 5.} {\it To calculate the optimal UAV's flight path as well as the optimal value for $L_{sc}$, it is better to first investigate the performance of the considered system versus $L_s$ since, it gives a better view in terms of finding the acceptable interval for $L_s$ and $L_d$.
		Then, to determine the optimal antenna pattern, we need to compute channel capacity by integrating on $\theta_{R1}$.}

	\section{Conclusion and Future Road Map}
	In this research, 
	taking into account the actual channel parameters such as the UAV vibrations, tracking error, real 3GPP antenna pattern, UAV's height and flight path, and considering the effect of physical obstacles, the optimal design of a relay system based on fixed wing UAV was investigated. In particular, the effects of the finding the optimal UAV's path as well as the optimal values for the number of antenna elements were studied.
	
	Although providing closed-form expressions for the outage probability and channel capacity allows faster and more accurate analysis, this is left as future work.
	Therefore, it is necessary to provide analytical expressions as a function of channel parameters.
	For longer link lengths, two- or multi-relay systems must be used and due to the variety of tunable parameters, the design of optimal parameters is much more complicated.
	Due to the continuous displacement of the drone relative to CN and RA, the use of adaptive methods to control power, transmission rate and antenna pattern can be attractive.
	Investigating the Doppler effect according to the speed and UAV's flight rout, studying the selection of the appropriate wavelength according to the physical conditions as well as weather conditions can be important issues in this subject.

	


\end{document}